\documentstyle[12pt]{book}
\begin{document}

\sloppy
\raggedbottom

\chapter* 
{THE LIMITS OF MATHEMATICS}
\markright
{The Limits of Mathematics}
\addcontentsline{toc}{chapter}
{The limits of mathematics}

\section*{G. J. Chaitin,
IBM Research Division,
P.~O. Box 704, Yorktown Heights, NY 10598,
chaitin@watson.ibm.com}
\section*{}

\section*{Introduction}

In a remarkable development, I have constructed a new definition for a
self-delimiting universal Turing machine (UTM) that is easy to program
and runs very quickly.  This provides a new foundation for algorithmic
information theory (AIT), which is the theory of the size in bits of
programs for self-delimiting UTM's.  Previously, AIT had an abstract
mathematical quality.  Now it is possible to write down executable
programs that embody the constructions in the proofs of theorems.  So
AIT goes from dealing with remote idealized mythical objects to being
a theory about practical down-to-earth gadgets that one can actually
play with and use.

This new self-delimiting UTM is implemented via software written in a
new version of LISP that I invented especially for this purpose.  This
LISP was designed by writing an interpreter for it in Mathematica that
was then translated into C\@. I have tested this software by running it on
IBM RS/6000 workstations with the AIX version of UNIX.

Using this new software and the latest theoretical ideas, it
is now possible to give a self-contained ``hands on'' mini-course
presenting very concretely my latest proofs of my two fundamental
information-theoretic incompleteness theorems.  The first of these
theorems states that an $N$-bit formal axiomatic system cannot enable
one to exhibit any specific object with program-size complexity
greater than $N+c$.  The second of these theorems states that an
$N$-bit formal axiomatic system cannot enable one to determine more
than $N+c'$ scattered bits of the halting probability $\Omega$.

Most people believe that anything that is true is true for a reason.
These theorems show that some things are true for no reason at all,
i.e., accidentally, or at random.

As is shown in this course, the algorithms considered in the
proofs of these two theorems are now easy to program and run, and by
looking at the size in bits of these programs one can actually, for
the first time, determine exact values for the constants $c$ and $c'$.

I used this approach and software in an intensive short course on the
limits of mathematics that I gave at the University of Maine in Orono
in the summer of 1994.  I also lectured on this material during a
stay at the Santa Fe Institute in the spring of 1995, and at a meeting
at the Black Sea University in Romania in the summer of 1995.  A
summary of the approach that I used on these three occasions will
appear under the title ``A new version of algorithmic information
theory'' in a forthcoming issue of the new magazine {\it Complexity,}
which has just been launched by the Santa Fe Institute and John Wiley
and Sons.  A less technical discussion of the basic ideas that are
involved ``How to run algorithmic information theory on a computer''
will also appear in {\it Complexity.}

After presenting this material at these three different places, it
became obvious to me that it is extremely difficult to understand it
in its original form.  So next time, at the Rovaniemi Institute of
Technology in the spring of 1996, I am going to use the new, more
understandable software in this report; everything has been redone in
an attempt to make it as easy to understand as possible.

For their stimulating invitations, I thank Prof.\ George Markowsky of
the University of Maine, Prof.\ Cristian Calude of the
University of Auckland, Prof.\ John Casti of the Santa Fe Institute,
and Prof.\ Veikko Ker\"anen of the Rovaniemi Institute of Technology.
And I am grateful to IBM for supporting my research for almost thirty
years, and to my current management chain at the IBM Research
Division, Dan Prener, Christos Georgiou, Eric Kronstadt, Jeff Jaffe,
and Jim McGroddy.

This report includes the LISP runs {\tt *.r} used to present the
information-theoretic incompleteness theorems of algorithmic
information theory.  This report does not include the software used to
produce these LISP runs.  To obtain the software for this course via
e-mail, please send requests to {\tt chaitin@watson.ibm.com}.

\section*{The New Idea}

Here is a quick summary of this new LISP, in which atoms can now either be
words or unsigned decimal integers.  First of all, comments are
written like this: {\tt [comment]}.  Each LISP primitive function has
a fixed number of arguments.  {\tt '} is {\tt QUOTE}, {\tt =} is {\tt
EQ}, and {\tt atom, car, cdr, cadr, caddr, cons} are provided with
their usual meaning.  We also have {\tt lambda, define, let, if} and
{\tt display} and {\tt eval}.  The notation {\tt "} indicates that an
S-expression with explicit parentheses follows, not what is usually
the case in this LISP, an M-expression, in which the parentheses for
each primitive function are implicit.  {\tt nil} denotes the empty
list {\tt ()}, and the logical truth values are {\tt true} and {\tt
false}.  For dealing with unsigned decimal integers we have \verb|+,
-, *, ^, <, >, <=, >=, base10-to-2, base2-to-10|.

So far this is fairly standard.  The new idea is this.  We define our
standard self-delimiting universal Turing machine as follows.  Its
program is in binary, and appears on a tape in the following form.
First comes a LISP expression, written in ASCII with 8 bits per
character, and terminated by an end-of-line character \verb|'\n'|.  The
TM reads in this LISP expression, and then evaluates it.  As it does
this, two new primitive functions {\tt read-bit} and {\tt read-exp}
with no arguments may be used to read more from the TM tape.  Both of
these functions explode if the tape is exhausted, killing the
computation.  {\tt read-bit} reads a single bit from the tape.
{\tt read-exp} reads in an entire LISP expression, in 8-bit character
chunks, until it reaches an end-of-line character \verb|'\n'|.

This is the only way that information on the TM tape may be accessed,
which forces it to be used in a self-delimiting fashion.  This is
because no algorithm can search for the end of the tape and then use
the length of the tape as data in the computation.  If an algorithm
attempts to read a bit that is not on the tape, the algorithm aborts.

How is information placed on the TM tape in the first place?  Well, in
the starting environment, the tape is empty and any attempt to read it
will give an error message.  To place information on the tape, one
must use the primitive function {\tt try} which tries to see if an
expression can be evaluated.

Consider the three arguments $\alpha$, $\beta$ and $\gamma$ of {\tt
try}.  The meaning of the first argument is as follows.  If $\alpha$
is {\tt no-time-limit}, then there is no depth limit.  Otherwise
$\alpha$ must be an unsigned decimal integer, and gives the depth
limit (limit on the nesting depth of function calls and
re-evaluations).  The second argument $\beta$ of {\tt try} is the
expression to be evaluated as long as the depth limit $\alpha$ is not
exceeded.  And the third argument $\gamma$ of {\tt try} is a list of
bits to be used as the TM tape.

The value $\nu$ returned by the primitive function {\tt try} is a
triple.  The first element of $\nu$ is {\tt success} if the evaluation
of $\beta$ was completed successfully, and the first element of $\nu$
is {\tt failure} if this was not the case.  The second element of
$\nu$ is {\tt out-of-data} if the evaluation of $\beta$ aborted
because an attempt was made to read a non-existent bit from the TM
tape.  The second element of $\nu$ is {\tt out-of-time} if evaluation
of $\beta$ aborted because the depth limit $\alpha$ was exceeded.
These are the only possible error flags, because this LISP is designed
with maximally permissive semantics.  If the computation $\beta$
terminated normally instead of aborting, the second element of $\nu$
will be the result produced by the computation $\beta$, i.e., its
value.  That's the second element of the list $\nu$ produced by the
{\tt try} primitive function.

The third element of the value $\nu$ is a list of all the arguments to
the primitive function {\tt display} that were encountered during the
evaluation of $\beta$.  More precisely, if {\tt display} was called
$N$ times during the evaluation of $\beta$, then $\nu$ will be a list
of $N$ elements.  The $N$ arguments of {\tt display} appear in $\nu$
in chronological order.  Thus {\tt try} can not only be used to
determine if a computation $\beta$ reads too much tape or goes on too
long (i.e., to greater depth than $\alpha$), but {\tt try} can also be
used to capture all the output that $\beta$ displayed as it went
along, whether the computation $\beta$ aborted or not.

In summary, all that one has to do to simulate a self-delimiting
universal Turing machine $U(p)$ running on the binary program $p$ is
to write
\begin{verbatim}
          try no-time-limit 'eval read-exp p
\end{verbatim}
This is an M-expression with parentheses omitted from primitive
functions.  (Recall that all primitive functions have a fixed number
of arguments.)  With the parentheses supplied, it becomes the
S-expression
\begin{verbatim}
          (try no-time-limit ('(eval(read-exp))) p)
\end{verbatim}
This says that one is to read a complete LISP S-expression from the TM
tape $p$ and then evaluate it without any time limit and using
whatever is left on the tape $p$.

Some more primitive functions have also been added.  The 2-argument
function {\tt append} denotes list concatenation, and the 1-argument
function {\tt bits} converts an S-expression into the list of the bits
in its ASCII character string representation.  These are used for
constructing the bit strings that are then put on the TM tape using
{\tt try}'s third argument $\gamma$.  We also provide the 1-argument
functions {\tt size} and {\tt length} that respectively give the
number of characters in an S-expression, and the number of elements in
a list.  Note that the functions {\tt append}, {\tt size} and {\tt
length} could be programmed rather than included as built-in primitive
functions, but it is extremely convenient and much much faster to
provide them built in.

Finally a new 1-argument identity function {\tt debug} with the
side-effect of outputting its argument is provided for debugging.
Output produced by {\tt debug} is invisible to the ``official'' {\tt
display} and {\tt try} output mechanism.  {\tt debug} is needed
because {\tt try} $\alpha$ $\beta$ $\gamma$ suppresses all output
$\theta$ produced within its depth-controlled evaluation of $\beta$.
Instead {\tt try} collects all output $\theta$ from within $\beta$ for
inclusion in the final value $\nu$ that {\tt try} returns, namely $\nu
= $ (success/failure, value of $\beta$, $\theta$).

\section*{Course Outline}

The course begins by explaining with examples my new LISP.  See {\tt
examples.r}.

Then the theory of LISP program-size complexity is developed a little
bit.  LISP program-size complexity is extremely simple and concrete.
In particular, it is easy to show that it is impossible to prove that
a self-contained LISP expression is elegant, i.e., that no smaller
expression has the same value.  To prove that an $N$-character LISP
expression is elegant requires a formal axiomatic system that itself
has at least LISP complexity $N-410$.  See {\tt godel.r}.

Next we define our standard self-delimiting universal Turing machine
$U(p)$ using
\begin{verbatim}
          cadr try no-time-limit 'eval read-exp p
\end{verbatim}
as explained in the previous chapter.

Next we show that
\[
   H(x,y) \le H(x) + H(y) + c
\]
with $c = 432.$
Here $H(\cdots)$ denotes the size in bits of the smallest program that
makes our standard universal Turing machine compute $\cdots$.  Thus
this inequality states that the information needed to compute the pair
$(x,y)$ is bounded by a constant $c$ plus the sum of the information
needed to compute $x$ and the information needed to compute $y$.
Consider
\begin{verbatim}
          cons eval read-exp
          cons eval read-exp
               nil
\end{verbatim}
This is an M-expression with parentheses omitted from primitive
functions.  With all the parentheses supplied, it becomes the S-expression
\begin{verbatim}
          (cons (eval (read-exp))
          (cons (eval (read-exp))
                nil))
\end{verbatim}
$c = 432$ is just 8 bits plus 8 times the size in characters of this LISP
S-expression.  See {\tt utm.r}.

Consider a binary string $x$ whose size is $|x|$ bits.  In {\tt
utm.r} we also show that
\[
   H(x) \le 2|x| + c
\]
and
\[
   H(x) \le |x| + H(|x|) + c'
\]
with $c = 1106$ and $c' = 1024$.  As before, the programs for doing
this are exhibited and run.

Next we turn to the self-delimiting program-size complexity $H(X)$ for
infinite r.e.\ sets $X$.  This is defined to be the size in bits of
the smallest LISP expression $\xi$ that executes forever without
halting and outputs the members of the r.e.\ set $X$ using the LISP
primitive {\tt display}, which is an identity function with the
side-effect of outputting the value of its argument.  Note that this
LISP expression $\xi$ is allowed to read additional bits or
expressions from the TM tape using the primitive functions {\tt
read-bit} and {\tt read-exp} if $\xi$ so desires.  But of course $\xi$
is charged for this; this adds to $\xi$'s program size.

It is in order to deal with such unending expressions $\xi$ that the
LISP primitive function for time-limited evaluation {\tt try} captures
all output from {\tt display} within its second argument $\beta$.

Now consider a formal axiomatic system $A$ of complexity $N$, i.e.,
with a set of theorems $T_A$ that considered as an r.e.\ set as above
has self-delimiting program-size complexity $H(T_A) = N$.  We show
that $A$ cannot enable us to exhibit a specific S-expression $s$ with
self-delimiting complexity $H(s)$ greater than $N+c$.  Here $c =
4872$.  See {\tt godel2.r}.

Next we show two different ways to calculate the halting probability
$\Omega$ of our standard self-delimiting universal Turing machine in
the limit from below.  See {\tt omega.r} and {\tt omega2.r}.  The
first way of doing this, {\tt omega.r}, is straight-forward.  The
second way to calculate $\Omega$, {\tt omega2.r}, uses a more clever
method.  Using the clever method as a subroutine, we show that if
$\Omega_N$ is the first $N$ bits of the fractional part of the
base-two real number $\Omega$, then
\[
   H(\Omega_N) > N - c
\]
with $c = 8000$.  Again this is done with a program that can actually
be run and whose size gives us a value for $c$.  See {\tt omega3.r}.

Consider again the formal axiomatic system $A$ with complexity $N$,
i.e., with self-delimiting program-size complexity $H(T_A) = N$.
Using the lower bound of $N-c$ on $H(\Omega_N)$ established in {\tt
omega3.r}, we show that $A$ cannot enable us to determine more than
the first $N+c'$ bits of $\Omega$.  Here $c' = 15328$.  In fact, we
show that $A$ cannot enable us to determine more than $N+c'$ bits of
$\Omega$ even if they are scattered and we leave gaps.  See {\tt
godel3.r}.

Last but not least, the philosophical implications of all this
should be discussed, especially the extent to which it tends to
justify experimental mathematics.  This would be along the lines of
the discussion in my talk transcript ``Randomness in arithmetic and
the decline and fall of reductionism in pure mathematics.''

This concludes our ``hands-on'' mini-course on the
information-theoretic limits of mathematics.

\section*{Bibliography}

Here is a useful collection of hand-outs for this course:

\begin{itemize}
\item[{[1]}] G. J. Chaitin, ``Randomness in arithmetic and the decline
and fall of reductionism in pure mathematics,'' in J. Cornwell, {\it
Nature's Imagination,} Oxford University Press, 1995, pp.\ 27--44.
\item[{[2]}] G. J. Chaitin, ``The Berry paradox,'' {\it Complexity\/}
1 (1995), pp.\ 26--30.
\item[{[3]}] G. J. Chaitin, ``A new version of algorithmic information
theory,'' {\it Complexity,} to appear.
\item[{[4]}] G. J. Chaitin, ``How to run algorithmic information
theory on a computer,'' {\it Complexity,} to appear.
\end{itemize}

\section*{examples.r}{\small\begin{verbatim}
LISP Interpreter Run

[ Test new lisp ]

' (ab c d)

expression  (' (ab c d))
value       (ab c d)

'(ab   cd  )

expression  (' (ab cd))
value       (ab cd)

car '(aa bb cc)

expression  (car (' (aa bb cc)))
value       aa

cdr '(aa bb cc)

expression  (cdr (' (aa bb cc)))
value       (bb cc)

cadr '(aa bb cc)

expression  (car (cdr (' (aa bb cc))))
value       bb

caddr '(aa bb cc)

expression  (car (cdr (cdr (' (aa bb cc)))))
value       cc

cons '(aa bb cc) '(dd ee ff)

expression  (cons (' (aa bb cc)) (' (dd ee ff)))
value       ((aa bb cc) dd ee ff)

car aa

expression  (car aa)
value       aa

cdr aa

expression  (cdr aa)
value       aa

cons aa bb

expression  (cons aa bb)
value       aa

nil

expression  nil
value       ()

cons aa nil

expression  (cons aa nil)
value       (aa)

("cons aa)

expression  (cons aa)
value       (aa)

("cons '(aa) '(bb) '(cc))

expression  (cons (' (aa)) (' (bb)) (' (cc)))
value       ((aa) bb)

let x a x

expression  ((' (lambda (x) x)) a)
value       a

x

expression  x
value       x

atom ' aa

expression  (atom (' aa))
value       true

atom '(aa)

expression  (atom (' (aa)))
value       false

if true x y

expression  (if true x y)
value       x

if false x y

expression  (if false x y)
value       y

if xxx x y

expression  (if xxx x y)
value       x

let (f x) if atom display x x (f car x)
 (f '(((a)b)c))

expression  ((' (lambda (f) (f (' (((a) b) c))))) (' (lambda (
            x) (if (atom (display x)) x (f (car x))))))
display     (((a) b) c)
display     ((a) b)
display     (a)
display     a
value       a

f

expression  f
value       f

append '(a b c) '(d e f)

expression  (append (' (a b c)) (' (d e f)))
value       (a b c d e f)

let (cat x y) if atom x y cons car x (cat cdr x y)
    (cat '(a b c) '(d e f))

expression  ((' (lambda (cat) (cat (' (a b c)) (' (d e f)))))
            (' (lambda (x y) (if (atom x) y (cons (car x) (cat
             (cdr x) y))))))
value       (a b c d e f)

cat

expression  cat
value       cat

define (cat x y) if atom x y cons car x (cat cdr x y)

define      cat
value       (lambda (x y) (if (atom x) y (cons (car x) (cat (c
            dr x) y))))

cat

expression  cat
value       (lambda (x y) (if (atom x) y (cons (car x) (cat (c
            dr x) y))))

(cat '(a b c) '(d e f))

expression  (cat (' (a b c)) (' (d e f)))
value       (a b c d e f)

define x (a b c)

define      x
value       (a b c)

cons x nil

expression  (cons x nil)
value       ((a b c))

define x (d e f)

define      x
value       (d e f)

cons x nil

expression  (cons x nil)
value       ((d e f))

length display bits ' a

expression  (length (display (bits (' a))))
display     (0 1 1 0 0 0 0 1 0 0 0 0 1 0 1 0)
value       16

length display bits ' abc

expression  (length (display (bits (' abc))))
display     (0 1 1 0 0 0 0 1 0 1 1 0 0 0 1 0 0 1 1 0 0 0 1 1 0
             0 0 0 1 0 1 0)
value       32

length display bits nil

expression  (length (display (bits nil)))
display     (0 0 1 0 1 0 0 0 0 0 1 0 1 0 0 1 0 0 0 0 1 0 1 0)
value       24

length display bits ' (a)

expression  (length (display (bits (' (a)))))
display     (0 0 1 0 1 0 0 0 0 1 1 0 0 0 0 1 0 0 1 0 1 0 0 1 0
             0 0 0 1 0 1 0)
value       32

size abc

expression  (size abc)
value       3

size ' ( a b c )

expression  (size (' (a b c)))
value       7

length ' ( a b c )

expression  (length (' (a b c)))
value       3

+ abc 15

expression  (+ abc 15)
value       15

+ '(abc) 15

expression  (+ (' (abc)) 15)
value       15

+ 10 15

expression  (+ 10 15)
value       25

- 10 15

expression  (- 10 15)
value       0

- 15 10

expression  (- 15 10)
value       5

* 10 15

expression  (* 10 15)
value       150

^ 10 15

expression  (^ 10 15)
value       1000000000000000

< 10 15

expression  (< 10 15)
value       true

< 10 10

expression  (< 10 10)
value       false

> 15 10

expression  (> 15 10)
value       true

> 10 10

expression  (> 10 10)
value       false

<= 15 10

expression  (<= 15 10)
value       false

<= 10 10

expression  (<= 10 10)
value       true

>= 10 15

expression  (>= 10 15)
value       false

>= 10 10

expression  (>= 10 10)
value       true

= 10 15

expression  (= 10 15)
value       false

= 10 10

expression  (= 10 10)
value       true

eval display '+ display 5 display 15

expression  (eval (display (' (+ (display 5) (display 15)))))
display     (+ (display 5) (display 15))
display     5
display     15
value       20

try 0 display '+ display 5 display 15 nil

expression  (try 0 (display (' (+ (display 5) (display 15))))
            nil)
display     (+ (display 5) (display 15))
value       (success 20 (5 15))

try 0 display '+ debug 5 debug 15 nil

expression  (try 0 (display (' (+ (debug 5) (debug 15)))) nil)
display     (+ (debug 5) (debug 15))
debug       5
debug       15
value       (success 20 ())

define five!
let (f x) if = 0 x 1 * display x (f - x 1)
    (f 5)

define      five!
value       ((' (lambda (f) (f 5))) (' (lambda (x) (if (= 0 x)
             1 (* (display x) (f (- x 1)))))))

eval five!

expression  (eval five!)
display     5
display     4
display     3
display     2
display     1
value       120

let (f x) if = 0 x 1 * x (f - x 1)
    (f 100)

expression  ((' (lambda (f) (f 100))) (' (lambda (x) (if (= 0
            x) 1 (* x (f (- x 1)))))))
value       93326215443944152681699238856266700490715968264381
            62146859296389521759999322991560894146397615651828
            62536979208272237582511852109168640000000000000000
            00000000

try 0 five! nil

expression  (try 0 five! nil)
value       (failure out-of-time ())

try 1 five! nil

expression  (try 1 five! nil)
value       (failure out-of-time ())

try 2 five! nil

expression  (try 2 five! nil)
value       (failure out-of-time (5))

try 3 five! nil

expression  (try 3 five! nil)
value       (failure out-of-time (5 4))

try 4 five! nil

expression  (try 4 five! nil)
value       (failure out-of-time (5 4 3))

try 5 five! nil

expression  (try 5 five! nil)
value       (failure out-of-time (5 4 3 2))

try 6 five! nil

expression  (try 6 five! nil)
value       (failure out-of-time (5 4 3 2 1))

try 7 five! nil

expression  (try 7 five! nil)
value       (success 120 (5 4 3 2 1))

try no-time-limit five! nil

expression  (try no-time-limit five! nil)
value       (success 120 (5 4 3 2 1))

define two*
 let (f x) if = 0 x nil
           cons * 2 display read-bit (f - x 1)
     (f 5)

define      two*
value       ((' (lambda (f) (f 5))) (' (lambda (x) (if (= 0 x)
             nil (cons (* 2 (display (read-bit))) (f (- x 1)))
            ))))

try 6 two* '(1 0 1 0 1)

expression  (try 6 two* (' (1 0 1 0 1)))
value       (failure out-of-time (1 0 1 0 1))

try 7 two* '(1 0 1 0 1)

expression  (try 7 two* (' (1 0 1 0 1)))
value       (success (2 0 2 0 2) (1 0 1 0 1))

try 7 two* '(1 0 1)

expression  (try 7 two* (' (1 0 1)))
value       (failure out-of-data (1 0 1))

try no-time-limit two* '(1 0 1)

expression  (try no-time-limit two* (' (1 0 1)))
value       (failure out-of-data (1 0 1))

try 18
'let (f x) if = 0 x nil
           cons * 2 display read-bit (f - x 1)
     (f 16)
bits 'a

expression  (try 18 (' ((' (lambda (f) (f 16))) (' (lambda (x)
             (if (= 0 x) nil (cons (* 2 (display (read-bit)))
            (f (- x 1)))))))) (bits (' a)))
value       (success (0 2 2 0 0 0 0 2 0 0 0 0 2 0 2 0) (0 1 1
            0 0 0 0 1 0 0 0 0 1 0 1 0))

base10-to-2 128

expression  (base10-to-2 128)
value       (1 0 0 0 0 0 0 0)

base10-to-2 256

expression  (base10-to-2 256)
value       (1 0 0 0 0 0 0 0 0)

base10-to-2 257

expression  (base10-to-2 257)
value       (1 0 0 0 0 0 0 0 1)

base2-to-10 '(1 1 1 1)

expression  (base2-to-10 (' (1 1 1 1)))
value       15

base2-to-10 '(1 0 0 0 0)

expression  (base2-to-10 (' (1 0 0 0 0)))
value       16

base2-to-10 '(1 0 0 0 1)

expression  (base2-to-10 (' (1 0 0 0 1)))
value       17

try 20
'cons abcdef try 10
'let (f n) (f display + n 1) (f 0) [infinite loop]
nil nil

expression  (try 20 (' (cons abcdef (try 10 (' ((' (lambda (f)
             (f 0))) (' (lambda (n) (f (display (+ n 1)))))))
            nil))) nil)
value       (success (abcdef failure out-of-time (1 2 3 4 5 6
            7 8 9)) ())

try 10
'cons abcdef try 20
'let (f n) (f display + n 1) (f 0) [infinite loop]
nil nil

expression  (try 10 (' (cons abcdef (try 20 (' ((' (lambda (f)
             (f 0))) (' (lambda (n) (f (display (+ n 1)))))))
            nil))) nil)
value       (failure out-of-time ())

try no-time-limit
'cons abcdef try 20
'let (f n) (f display + n 1) (f 0) [infinite loop]
nil nil

expression  (try no-time-limit (' (cons abcdef (try 20 (' (('
            (lambda (f) (f 0))) (' (lambda (n) (f (display (+
            n 1))))))) nil))) nil)
value       (success (abcdef failure out-of-time (1 2 3 4 5 6
            7 8 9 10 11 12 13 14 15 16 17 18 19)) ())

try 10
'cons abcdef try no-time-limit
'let (f n) (f display + n 1) (f 0) [infinite loop]
nil nil

expression  (try 10 (' (cons abcdef (try no-time-limit (' (('
            (lambda (f) (f 0))) (' (lambda (n) (f (display (+
            n 1))))))) nil))) nil)
value       (failure out-of-time ())

read-bit

expression  (read-bit)
value       out-of-data

read-exp

expression  (read-exp)
value       out-of-data

bits '(abc def)

expression  (bits (' (abc def)))
value       (0 0 1 0 1 0 0 0 0 1 1 0 0 0 0 1 0 1 1 0 0 0 1 0 0
             1 1 0 0 0 1 1 0 0 1 0 0 0 0 0 0 1 1 0 0 1 0 0 0 1
             1 0 0 1 0 1 0 1 1 0 0 1 1 0 0 0 1 0 1 0 0 1 0 0 0
             0 1 0 1 0)

try no-time-limit 'read-exp bits '(abc def)

expression  (try no-time-limit (' (read-exp)) (bits (' (abc de
            f))))
value       (success (abc def) ())

bits 'abc

expression  (bits (' abc))
value       (0 1 1 0 0 0 0 1 0 1 1 0 0 0 1 0 0 1 1 0 0 0 1 1 0
             0 0 0 1 0 1 0)

try 0 'read-bit nil

expression  (try 0 (' (read-bit)) nil)
value       (failure out-of-data ())

try 0 'read-exp nil

expression  (try 0 (' (read-exp)) nil)
value       (failure out-of-data ())

try 0 'read-exp bits 'abc

expression  (try 0 (' (read-exp)) (bits (' abc)))
value       (success abc ())

try 0 'cons read-exp cons read-bit nil bits 'abc

expression  (try 0 (' (cons (read-exp) (cons (read-bit) nil)))
             (bits (' abc)))
value       (failure out-of-data ())

try 0 'cons read-exp cons read-bit nil append bits 'abc '(0)

expression  (try 0 (' (cons (read-exp) (cons (read-bit) nil)))
             (append (bits (' abc)) (' (0))))
value       (success (abc 0) ())

try 0 'cons read-exp cons read-bit nil append bits 'abc '(1)

expression  (try 0 (' (cons (read-exp) (cons (read-bit) nil)))
             (append (bits (' abc)) (' (1))))
value       (success (abc 1) ())

try 0 'read-exp bits '(a b)

expression  (try 0 (' (read-exp)) (bits (' (a b))))
value       (success (a b) ())

try 0 'cons read-exp cons read-bit nil bits '(a b)

expression  (try 0 (' (cons (read-exp) (cons (read-bit) nil)))
             (bits (' (a b))))
value       (failure out-of-data ())

try 0 'cons read-exp cons read-exp nil bits '(a b)

expression  (try 0 (' (cons (read-exp) (cons (read-exp) nil)))
             (bits (' (a b))))
value       (failure out-of-data ())

try 0 'read-exp bits '(abc(def ghi)j)

expression  (try 0 (' (read-exp)) (bits (' (abc (def ghi) j)))
            )
value       (success (abc (def ghi) j) ())

try 0 'read-exp '(1 1 1 1) [character is incomplete]

expression  (try 0 (' (read-exp)) (' (1 1 1 1)))
value       (failure out-of-data ())

try 0 'read-exp
      '(0 0 0 0 1 0 1 0) [nothing in record; only \n]

expression  (try 0 (' (read-exp)) (' (0 0 0 0 1 0 1 0)))
value       (success () ())

try 0 'cons read-exp cons read-exp nil
      append bits '(a b c) bits '(d e f)

expression  (try 0 (' (cons (read-exp) (cons (read-exp) nil)))
             (append (bits (' (a b c))) (bits (' (d e f)))))
value       (success ((a b c) (d e f)) ())

try 0 'read-exp '(1 1 1 1  1 1 1 1
                  0 0 0 0  1 0 1 0) [invalid character]

expression  (try 0 (' (read-exp)) (' (1 1 1 1 1 1 1 1 0 0 0 0
            1 0 1 0)))
value       (success () ())

= 0003 3

expression  (= 3 3)
value       true

000099

expression  99
value       99

x

expression  x
value       (d e f)

let x b x

expression  ((' (lambda (x) x)) b)
value       b

x

expression  x
value       (d e f)

let 99 45 99

expression  ((' (lambda (99) 99)) 45)
value       99

End of LISP Run

Elapsed time is 13 seconds.
\end{verbatim}
}\section*{godel.r}{\small\begin{verbatim}
LISP Interpreter Run

[[[
    Show that a formal system of lisp complexity
    H_lisp (FAS) = N cannot enable us to exhibit
    an elegant S-expression of size greater than N + 410.
    An elegant lisp expression is one with the property
    that no smaller S-expression has the same value.
    Setting: formal axiomatic system is never-ending
    lisp expression that displays elegant S-expressions.
]]]

[Here is the key expression.]

define expression

let (examine x)
    if atom x  false
    if < n size car x  car x
    (examine cdr x)

let fas 'display ^ 10 430 [insert FAS here preceeded by ']

let n + 410 size fas

let t 0

let (loop)
  let v try t fas nil
  let s (examine caddr v)
  if s eval s
  if = success car v failure
  let t + t 1
  (loop)

(loop)

define      expression
value       ((' (lambda (examine) ((' (lambda (fas) ((' (lambd
            a (n) ((' (lambda (t) ((' (lambda (loop) (loop)))
            (' (lambda () ((' (lambda (v) ((' (lambda (s) (if
            s (eval s) (if (= success (car v)) failure ((' (la
            mbda (t) (loop))) (+ t 1)))))) (examine (car (cdr
            (cdr v))))))) (try t fas nil))))))) 0))) (+ 410 (s
            ize fas))))) (' (display (^ 10 430)))))) (' (lambd
            a (x) (if (atom x) false (if (< n (size (car x)))
            (car x) (examine (cdr x)))))))

[Size expression.]
size expression

expression  (size expression)
value       430

[Run expression & show that it knows its own size
 and can find something bigger than it is.]
eval expression

expression  (eval expression)
value       10000000000000000000000000000000000000000000000000
            00000000000000000000000000000000000000000000000000
            00000000000000000000000000000000000000000000000000
            00000000000000000000000000000000000000000000000000
            00000000000000000000000000000000000000000000000000
            00000000000000000000000000000000000000000000000000
            00000000000000000000000000000000000000000000000000
            00000000000000000000000000000000000000000000000000
            0000000000000000000000000000000

[Here it fails to find anything bigger than it is.]

let (examine x)
    if atom x  false
    if < n size car x  car x
    (examine cdr x)

let fas 'display ^ 10 429 [insert FAS here preceeded by ']

let n + 410 size fas

let t 0

let (loop)
  let v try t fas nil
  let s (examine caddr v)
  if s eval s
  if = success car v failure
  let t + t 1
  (loop)

(loop)

expression  ((' (lambda (examine) ((' (lambda (fas) ((' (lambd
            a (n) ((' (lambda (t) ((' (lambda (loop) (loop)))
            (' (lambda () ((' (lambda (v) ((' (lambda (s) (if
            s (eval s) (if (= success (car v)) failure ((' (la
            mbda (t) (loop))) (+ t 1)))))) (examine (car (cdr
            (cdr v))))))) (try t fas nil))))))) 0))) (+ 410 (s
            ize fas))))) (' (display (^ 10 429)))))) (' (lambd
            a (x) (if (atom x) false (if (< n (size (car x)))
            (car x) (examine (cdr x)))))))
value       failure

End of LISP Run

Elapsed time is 2 seconds.
\end{verbatim}
}\section*{utm.r}{\small\begin{verbatim}
LISP Interpreter Run

[[[
 First steps with my new construction for
 a self-delimiting universal Turing machine.
 We show that
    H(x,y) <= H(x) + H(y) + c
 and determine c.
 Consider a bit string x of length |x|.
 We also show that
    H(x) <= 2|x| + c
 and that
    H(x) <= |x| + H(the binary string for |x|) + c
 and determine both these c's.
]]]

[
 Here is the self-delimiting universal Turing machine!
]
define (U p) cadr try no-time-limit 'eval read-exp p

define      U
value       (lambda (p) (car (cdr (try no-time-limit (' (eval
            (read-exp))) p))))

(U bits 'cons x cons y cons z nil)

expression  (U (bits (' (cons x (cons y (cons z nil))))))
value       (x y z)

(U append bits 'cons a debug read-exp
          bits '(b c d)
)

expression  (U (append (bits (' (cons a (debug (read-exp)))))
            (bits (' (b c d)))))
debug       (b c d)
value       (a b c d)

[
 The length of alpha in bits is the
 constant c in H(x) <= 2|x| + 2 + c.
]
define alpha
let (loop) let x read-bit
           let y read-bit
           if = x y
              cons x (loop)
              nil
(loop)

define      alpha
value       ((' (lambda (loop) (loop))) (' (lambda () ((' (lam
            bda (x) ((' (lambda (y) (if (= x y) (cons x (loop)
            ) nil))) (read-bit)))) (read-bit)))))

length bits alpha

expression  (length (bits alpha))
value       1104

(U
 append
   bits alpha
   '(0 0 1 1 0 0 1 1 0 1)
)

expression  (U (append (bits alpha) (' (0 0 1 1 0 0 1 1 0 1)))
            )
value       (0 1 0 1)

(U
 append
   bits alpha
   '(0 0 1 1 0 0 1 1 0 0)
)

expression  (U (append (bits alpha) (' (0 0 1 1 0 0 1 1 0 0)))
            )
value       out-of-data

[
 The length of beta in bits is the
 constant c in H(x,y) <= H(x) + H(y) + c.
]
define beta
cons eval read-exp
cons eval read-exp
     nil

define      beta
value       (cons (eval (read-exp)) (cons (eval (read-exp)) ni
            l))

length bits beta

expression  (length (bits beta))
value       432

(U
 append
   bits  beta
 append
   bits 'cons a cons b cons c nil
   bits 'cons x cons y cons z nil
)

expression  (U (append (bits beta) (append (bits (' (cons a (c
            ons b (cons c nil))))) (bits (' (cons x (cons y (c
            ons z nil))))))))
value       ((a b c) (x y z))

(U
 append
   bits beta
 append
   append bits alpha '(0 0 1 1 0 0 1 1 0 1)
   append bits alpha '(1 1 0 0 1 1 0 0 1 0)
)

expression  (U (append (bits beta) (append (append (bits alpha
            ) (' (0 0 1 1 0 0 1 1 0 1))) (append (bits alpha)
            (' (1 1 0 0 1 1 0 0 1 0))))))
value       ((0 1 0 1) (1 0 1 0))

[
 The length of gamma in bits is the
 constant c in H(x) <= |x| + H(|x|) + c
]
define gamma
let (loop k)
   if = 0 k nil
   cons read-bit (loop - k 1)
(loop base2-to-10 eval read-exp)

define      gamma
value       ((' (lambda (loop) (loop (base2-to-10 (eval (read-
            exp)))))) (' (lambda (k) (if (= 0 k) nil (cons (re
            ad-bit) (loop (- k 1)))))))

length bits gamma

expression  (length (bits gamma))
value       1024

(U
 append
   bits gamma
 append
   [Arbitrary program for U to compute number of bits]
   bits' '(1 0 0 0)
   [That many bits of data]
   '(0 0 0 0  0 0 0 1)
)

expression  (U (append (bits gamma) (append (bits (' (' (1 0 0
             0)))) (' (0 0 0 0 0 0 0 1)))))
value       (0 0 0 0 0 0 0 1)

End of LISP Run

Elapsed time is 19 seconds.
\end{verbatim}
}\section*{godel2.r}{\small\begin{verbatim}
LISP Interpreter Run

[[[
 Show that a formal system of complexity N
 can't prove that a specific object has
 complexity > N + 4872.
 Formal system is a never halting lisp expression
 that output pairs (lisp object, lower bound
 on its complexity).  E.g., (x 4) means
 that x has complexity H(x) greater than or equal to 4.
]]]

[Here is the prefix.]

define pi

let (examine pairs)
    if atom pairs false
    if < n cadr car pairs
       car pairs
       (examine cdr pairs)

let t 0

let fas nil

let (loop)
  let v try t 'eval read-exp fas
  let n + 4872 length fas
  let p (examine caddr v)
  if p car p
  if = car v success failure
  if = cadr v out-of-data
     let fas append fas cons read-bit nil
     (loop)
  if = cadr v out-of-time
     let t + t 1
     (loop)
  unexpected-condition

(loop)

define      pi
value       ((' (lambda (examine) ((' (lambda (t) ((' (lambda
            (fas) ((' (lambda (loop) (loop))) (' (lambda () ((
            ' (lambda (v) ((' (lambda (n) ((' (lambda (p) (if
            p (car p) (if (= (car v) success) failure (if (= (
            car (cdr v)) out-of-data) ((' (lambda (fas) (loop)
            )) (append fas (cons (read-bit) nil))) (if (= (car
             (cdr v)) out-of-time) ((' (lambda (t) (loop))) (+
             t 1)) unexpected-condition)))))) (examine (car (c
            dr (cdr v))))))) (+ 4872 (length fas))))) (try t (
            ' (eval (read-exp))) fas))))))) nil))) 0))) (' (la
            mbda (pairs) (if (atom pairs) false (if (< n (car
            (cdr (car pairs)))) (car pairs) (examine (cdr pair
            s)))))))

[Size pi.]
length bits pi

expression  (length (bits pi))
value       4872

[Size pi + fas.]
length
append bits pi
       bits 'display '(xyz 9999)

expression  (length (append (bits pi) (bits (' (display (' (xy
            z 9999)))))))
value       5072

[Here pi finds something suitable.]

cadr try no-time-limit 'eval read-exp
append bits pi
       bits 'display '(xyz 5073)

expression  (car (cdr (try no-time-limit (' (eval (read-exp)))
             (append (bits pi) (bits (' (display (' (xyz 5073)
            ))))))))
value       xyz

[Here pi doesn't find anything suitable.]

cadr try no-time-limit 'eval read-exp
append bits pi
       bits 'display '(xyz 5072)

expression  (car (cdr (try no-time-limit (' (eval (read-exp)))
             (append (bits pi) (bits (' (display (' (xyz 5072)
            ))))))))
value       failure

End of LISP Run

Elapsed time is 153 seconds.
\end{verbatim}
}\section*{omega.r}{\small\begin{verbatim}
LISP Interpreter Run

[[[[ Omega in the limit from below! ]]]]

define (all-bit-strings-of-size k)
    if = 0 k '(())
    (extend-by-one-bit (all-bit-strings-of-size - k 1))

define      all-bit-strings-of-size
value       (lambda (k) (if (= 0 k) (' (())) (extend-by-one-bi
            t (all-bit-strings-of-size (- k 1)))))

define (extend-by-one-bit x)
    if atom x nil
    cons append car x '(0)
    cons append car x '(1)
    (extend-by-one-bit cdr x)

define      extend-by-one-bit
value       (lambda (x) (if (atom x) nil (cons (append (car x)
             (' (0))) (cons (append (car x) (' (1))) (extend-b
            y-one-bit (cdr x))))))

define (count-halt p)
    if atom p 0
    +
    if = success car try t 'eval read-exp car p
       1 0
    (count-halt cdr p)

define      count-halt
value       (lambda (p) (if (atom p) 0 (+ (if (= success (car
            (try t (' (eval (read-exp))) (car p)))) 1 0) (coun
            t-halt (cdr p)))))

define (omega t) cons (count-halt (all-bit-strings-of-size t))
                 cons /
                 cons ^ 2 t
                      nil

define      omega
value       (lambda (t) (cons (count-halt (all-bit-strings-of-
            size t)) (cons / (cons (^ 2 t) nil))))

(omega 0)

expression  (omega 0)
value       (0 / 1)

(omega 1)

expression  (omega 1)
value       (0 / 2)

(omega 2)

expression  (omega 2)
value       (0 / 4)

(omega 3)

expression  (omega 3)
value       (0 / 8)

(omega 8)

expression  (omega 8)
value       (1 / 256)

End of LISP Run

Elapsed time is 38 seconds.
\end{verbatim}
}\section*{omega2.r}{\small\begin{verbatim}
LISP Interpreter Run

[[[[ Omega in the limit from below! ]]]]

define (count-halt prefix bits-left-to-extend)
    if = bits-left-to-extend 0
    if = success car try t 'eval read-exp prefix
       1 0
    + (count-halt append prefix '(0) - bits-left-to-extend 1)
      (count-halt append prefix '(1) - bits-left-to-extend 1)

define      count-halt
value       (lambda (prefix bits-left-to-extend) (if (= bits-l
            eft-to-extend 0) (if (= success (car (try t (' (ev
            al (read-exp))) prefix))) 1 0) (+ (count-halt (app
            end prefix (' (0))) (- bits-left-to-extend 1)) (co
            unt-halt (append prefix (' (1))) (- bits-left-to-e
            xtend 1)))))

define (omega t) cons (count-halt nil t)
                 cons /
                 cons ^ 2 t
                      nil

define      omega
value       (lambda (t) (cons (count-halt nil t) (cons / (cons
             (^ 2 t) nil))))

(omega 0)

expression  (omega 0)
value       (0 / 1)

(omega 1)

expression  (omega 1)
value       (0 / 2)

(omega 2)

expression  (omega 2)
value       (0 / 4)

(omega 3)

expression  (omega 3)
value       (0 / 8)

(omega 8)

expression  (omega 8)
value       (1 / 256)

End of LISP Run

Elapsed time is 33 seconds.
\end{verbatim}
}\section*{omega3.r}{\small\begin{verbatim}
LISP Interpreter Run

[[[
 Show that
    H(Omega_n) > n - 8000.
 Omega_n is the first n bits of Omega,
 where we choose
    Omega = xxx0111111...
 instead of
    Omega = xxx1000000...
 if necessary.
]]]

[Here is the prefix.]

define pi

let (count-halt prefix bits-left-to-extend)
    if = bits-left-to-extend 0
    if = success car try t 'eval read-exp prefix
       1 0
    + (count-halt append prefix '(0) - bits-left-to-extend 1)
      (count-halt append prefix '(1) - bits-left-to-extend 1)

let (omega t) cons (count-halt nil t)
              cons /
              cons ^ 2 t
                   nil

let w eval read-exp

let n length w

let w cons base2-to-10 w
      cons /
      cons ^ 2 n
           nil

let (loop t)
  if (<=rat w (omega t))
     (big nil n)
     (loop + t 1)

let (<=rat x y)
    <= * car x caddr y * caddr x car y

let (big prefix bits-left-to-add)
 if = 0 bits-left-to-add
 cons cadr try t 'eval read-exp prefix
      nil
 append (big append prefix '(0) - bits-left-to-add 1)
        (big append prefix '(1) - bits-left-to-add 1)

(loop 0)

define      pi
value       ((' (lambda (count-halt) ((' (lambda (omega) ((' (
            lambda (w) ((' (lambda (n) ((' (lambda (w) ((' (la
            mbda (loop) ((' (lambda (<=rat) ((' (lambda (big)
            (loop 0))) (' (lambda (prefix bits-left-to-add) (i
            f (= 0 bits-left-to-add) (cons (car (cdr (try t ('
             (eval (read-exp))) prefix))) nil) (append (big (a
            ppend prefix (' (0))) (- bits-left-to-add 1)) (big
             (append prefix (' (1))) (- bits-left-to-add 1))))
            ))))) (' (lambda (x y) (<= (* (car x) (car (cdr (c
            dr y)))) (* (car (cdr (cdr x))) (car y)))))))) ('
            (lambda (t) (if (<=rat w (omega t)) (big nil n) (l
            oop (+ t 1)))))))) (cons (base2-to-10 w) (cons / (
            cons (^ 2 n) nil)))))) (length w)))) (eval (read-e
            xp))))) (' (lambda (t) (cons (count-halt nil t) (c
            ons / (cons (^ 2 t) nil)))))))) (' (lambda (prefix
             bits-left-to-extend) (if (= bits-left-to-extend 0
            ) (if (= success (car (try t (' (eval (read-exp)))
             prefix))) 1 0) (+ (count-halt (append prefix (' (
            0))) (- bits-left-to-extend 1)) (count-halt (appen
            d prefix (' (1))) (- bits-left-to-extend 1)))))))

[Run pi.]
cadr try no-time-limit 'eval read-exp
append bits pi
       bits '
      [Program to compute first n = 8 bits of Omega]
            '(0 0 0 0  0 0 0 1)

expression  (car (cdr (try no-time-limit (' (eval (read-exp)))
             (append (bits pi) (bits (' (' (0 0 0 0 0 0 0 1)))
            )))))
value       (out-of-data out-of-data out-of-data out-of-data o
            ut-of-data out-of-data out-of-data out-of-data out
            -of-data out-of-data () out-of-data out-of-data ou
            t-of-data out-of-data out-of-data out-of-data out-
            of-data out-of-data out-of-data out-of-data out-of
            -data out-of-data out-of-data out-of-data out-of-d
            ata out-of-data out-of-data out-of-data out-of-dat
            a out-of-data out-of-data out-of-data out-of-data
            out-of-data out-of-data out-of-data out-of-data ou
            t-of-data out-of-data out-of-data out-of-data out-
            of-data out-of-data out-of-data out-of-data out-of
            -data out-of-data out-of-data out-of-data out-of-d
            ata out-of-data out-of-data out-of-data out-of-dat
            a out-of-data out-of-data out-of-data out-of-data
            out-of-data out-of-data out-of-data out-of-data ou
            t-of-data out-of-data out-of-data out-of-data out-
            of-data out-of-data out-of-data out-of-data out-of
            -data out-of-data out-of-data out-of-data out-of-d
            ata out-of-data out-of-data out-of-data out-of-dat
            a out-of-data out-of-data out-of-data out-of-data
            out-of-data out-of-data out-of-data out-of-data ou
            t-of-data out-of-data out-of-data out-of-data out-
            of-data out-of-data out-of-data out-of-data out-of
            -data out-of-data out-of-data out-of-data out-of-d
            ata out-of-data out-of-data out-of-data out-of-dat
            a out-of-data out-of-data out-of-data out-of-data
            out-of-data out-of-data out-of-data out-of-data ou
            t-of-data out-of-data out-of-data out-of-data out-
            of-data out-of-data out-of-data out-of-data out-of
            -data out-of-data out-of-data out-of-data out-of-d
            ata out-of-data out-of-data out-of-data out-of-dat
            a out-of-data out-of-data out-of-data out-of-data
            out-of-data out-of-data out-of-data out-of-data ou
            t-of-data out-of-data out-of-data out-of-data out-
            of-data out-of-data out-of-data out-of-data out-of
            -data out-of-data out-of-data out-of-data out-of-d
            ata out-of-data out-of-data out-of-data out-of-dat
            a out-of-data out-of-data out-of-data out-of-data
            out-of-data out-of-data out-of-data out-of-data ou
            t-of-data out-of-data out-of-data out-of-data out-
            of-data out-of-data out-of-data out-of-data out-of
            -data out-of-data out-of-data out-of-data out-of-d
            ata out-of-data out-of-data out-of-data out-of-dat
            a out-of-data out-of-data out-of-data out-of-data
            out-of-data out-of-data out-of-data out-of-data ou
            t-of-data out-of-data out-of-data out-of-data out-
            of-data out-of-data out-of-data out-of-data out-of
            -data out-of-data out-of-data out-of-data out-of-d
            ata out-of-data out-of-data out-of-data out-of-dat
            a out-of-data out-of-data out-of-data out-of-data
            out-of-data out-of-data out-of-data out-of-data ou
            t-of-data out-of-data out-of-data out-of-data out-
            of-data out-of-data out-of-data out-of-data out-of
            -data out-of-data out-of-data out-of-data out-of-d
            ata out-of-data out-of-data out-of-data out-of-dat
            a out-of-data out-of-data out-of-data out-of-data
            out-of-data out-of-data out-of-data out-of-data ou
            t-of-data out-of-data out-of-data out-of-data out-
            of-data out-of-data out-of-data out-of-data out-of
            -data out-of-data out-of-data out-of-data out-of-d
            ata out-of-data out-of-data out-of-data out-of-dat
            a out-of-data)

[Size pi.]
length bits pi

expression  (length (bits pi))
value       8000

End of LISP Run

Elapsed time is 148 seconds.
\end{verbatim}
}\section*{godel3.r}{\small\begin{verbatim}
LISP Interpreter Run

[[[
 Show that a formal system of complexity N
 can't determine more than N + 8000 + 7328
 = N + 15328 bits of Omega.
 Formal system is a never halting lisp expression
 that outputs lists of the form (1 0 X 0 X X X X 1 0).
 This stands for the fractional part of Omega,
 and means that these 0,1 bits of Omega are known.
 X stands for an unknown bit.
]]]

[Here is the prefix.]

define pi

let (number-of-bits-determined w)
    if atom w 0
    + (number-of-bits-determined cdr w)
      if = X car w
         0
         1

let (supply-missing-bits w)
    if atom w nil
    cons if = X car w
            read-bit
            car w
    (supply-missing-bits cdr w)

let (examine w)
    if atom w false
   [if < n (number-of-bits-determined car w)]
   [   Change n to 1 here so will succeed.  ]
    if < 1 (number-of-bits-determined car w)
       car w
       (examine cdr w)

let t 0

let fas nil

let (loop)
  let v try t 'eval read-exp fas
  let n + 8000 + 7328 length fas
  let w (examine caddr v)
  if w (supply-missing-bits w)
  if = car v success failure
  if = cadr v out-of-data
     let fas append fas cons read-bit nil
     (loop)
  if = cadr v out-of-time
     let t + t 1
     (loop)
  unexpected-condition

(loop)

define      pi
value       ((' (lambda (number-of-bits-determined) ((' (lambd
            a (supply-missing-bits) ((' (lambda (examine) (('
            (lambda (t) ((' (lambda (fas) ((' (lambda (loop) (
            loop))) (' (lambda () ((' (lambda (v) ((' (lambda
            (n) ((' (lambda (w) (if w (supply-missing-bits w)
            (if (= (car v) success) failure (if (= (car (cdr v
            )) out-of-data) ((' (lambda (fas) (loop))) (append
             fas (cons (read-bit) nil))) (if (= (car (cdr v))
            out-of-time) ((' (lambda (t) (loop))) (+ t 1)) une
            xpected-condition)))))) (examine (car (cdr (cdr v)
            )))))) (+ 8000 (+ 7328 (length fas)))))) (try t ('
             (eval (read-exp))) fas))))))) nil))) 0))) (' (lam
            bda (w) (if (atom w) false (if (< 1 (number-of-bit
            s-determined (car w))) (car w) (examine (cdr w))))
            ))))) (' (lambda (w) (if (atom w) nil (cons (if (=
             X (car w)) (read-bit) (car w)) (supply-missing-bi
            ts (cdr w))))))))) (' (lambda (w) (if (atom w) 0 (
            + (number-of-bits-determined (cdr w)) (if (= X (ca
            r w)) 0 1))))))

[Size pi.]
length bits pi

expression  (length (bits pi))
value       7328

[Run pi.]

cadr try no-time-limit 'eval read-exp
append bits pi
append [Toy formal system with only one theorem.]
       bits 'display '(1 X 0)
       [Missing bit of omega that is needed.]
       '(1)

expression  (car (cdr (try no-time-limit (' (eval (read-exp)))
             (append (bits pi) (append (bits (' (display (' (1
             X 0))))) (' (1)))))))
value       (1 1 0)

End of LISP Run

Elapsed time is 94 seconds.
\end{verbatim}
}

\end{document}